\documentclass[twoside]{article}%
\usepackage{hyperref}
\usepackage{graphicx} 
\usepackage{multicol} \usepackage{rotating} 
\usepackage{floatflt}
\def\sqr#1#2{{\vbox{\hrule height.#2pt\hbox{\vrule width.#2pt
height#1pt \kern#1pt \vrule width.#2pt}\hrule height.#2pt}}}

\newcommand{\ap}[1]{\ifmmode^{\rm #1}\else $^{\rm #1}$\fi}

\newcommand{\ds}{\partial\!\!\!\raisebox{2pt}[0pt][0pt]{$\scriptstyle/$}}

\def\diag{\mathop{\rm diag}}

\def\One{\hbox{1\kern-.24em I}}

\def\circa#1{\raise.3ex\hbox{$#1$\kern-.75em\lower1ex\hbox{$\sim$}}}

\def\mathletters{\refstepcounter{equation}%
  \edef\@savedequation{\the\c@equation}%
  \@stequation=\expandafter{\theequation}%   %only want \theequation
  \edef\@savedtheequation{\the\@stequation}% %expanded once
  \edef\oldtheequation{\theequation}%
  \setcounter{equation}{0}%
  \def\theequation{\oldtheequation\alph{equation}}}

\def\endmathletters{%
  \setcounter{equation}{\@savedequation}%
  \@stequation=\expandafter{\@savedtheequation}%
  \edef\theequation{\the\@stequation}%
  \global\@ignoretrue}
\newcommand{\erf}{\hbox{erf}\,}

\makeatletter
\newcounter{alphaequation}[equation]
\def\thealphaequation{\theequation\hbox to
0.6em{\hfil\alph{alphaequation}\hfil}}
\def\eqnsystem#1{
\def\@eqnnum{{\rm (\thealphaequation)}}
\def\@@eqncr{\let\@tempa\relax \ifcase\@eqcnt \def\@tempa{& & &} \or
  \def\@tempa{& &}\or \def\@tempa{&}\fi\@tempa
  \if@eqnsw\@eqnnum\refstepcounter{alphaequation}\fi
\global\@eqnswtrue\global\@eqcnt=0\cr}
\refstepcounter{equation} \let\@currentlabel\theequation \def\@tempb{#1}
\ifx\@tempb\empty\else\label{#1}\fi
\refstepcounter{alphaequation}
\let\@currentlabel\thealphaequation
\global\@eqnswtrue\global\@eqcnt=0 \tabskip\@centering\let\\=\@eqncr
$$\halign to \displaywidth\bgroup \@eqnsel\hskip\@centering
$\displaystyle\tabskip\z@{##}$&\global\@eqcnt\@ne
\hskip2\arraycolsep\hfil${##}$\hfil& \global\@eqcnt\tw@\hskip2\arraycolsep
$\displaystyle\tabskip\z@{##}$\hfil
\tabskip\@centering&\llap{##}\tabskip\z@\cr}

\def\endeqnsystem{\@@eqncr\egroup$$\global\@ignoretrue} \makeatother

\def\MeV{\,{\rm MeV}}
\def\circa#1{\,\raise.3ex\hbox{$#1$\kern-.75em\lower1ex\hbox{$\sim$}}\,}

\newcommand{\riga}[1]{\noalign{\hbox{\parbox{\textwidth}{#1}}}\nonumber}
\newcount\Mac  \Mac=1  % devo mettere Mac=1 se sto lavorando sul file Mac
\newcommand{\ifMac}[2]{\ifnum\Mac=1 #1 \else #2 \fi}
\def\putps(#1,#2)(#3,#4)#5#6{\ifnum\Mac=1 \put(#1,#2){\special{picture #5}}
\else  \put(#3,#4){\special{epsfile=#6}} \fi}

\def\One{\hbox{1\kern-.24em I}}

\newcommand{\eV}{\,{\rm eV}}
\newcommand{\GeV}{\,{\rm GeV}}
\newcommand{\TeV}{\,{\rm TeV}}

\newcommand{\hc}{\hbox{h.c.}}

\newcommand{\eq}[1]{~{\rm (\ref{eq:#1})}}

\newcommand{\fig}[1]{~{\rm \ref{fig:#1}}}
\newcommand{\tab}[1]{~{\ref{tab:#1}}}

\def\Red{}
\def\Black{}
\def\Blue{}

\newcommand{\NP}{Nucl. Phys.}

\newcommand{\PRL}{Phys. Rev. Lett.}
\newcommand{\PL}{Phys. Lett.}
\newcommand{\PR}{Phys. Rev.}
\makeatletter
\def\art{\@ifnextchar[{\eart}{\oart}}
\def\eart[#1]#2#3#4#5#6{{\rm #2}, {\em #3 \rm #4} {\rm (#6) #5} ({\em #1})}
\def\hepart[#1]#2{{\rm #2, \em#1}}
\newcommand{\oart}[5]{{\rm #1}, {\em #2 \rm #3} {\rm (#5) #4}}

\newcommand{\biriga}[1]{\vbox{\hbox to\hsize{\hrulefill}
\kern-.5\baselineskip \hbox to\hsize{\hrulefill\phantom{ #1 }\hrulefill}
\kern-.5\baselineskip \hbox to\hsize{\hrulefill   \hbox{ #1 }\hrulefill}
\kern-.5\baselineskip \hbox to\hsize{\hrulefill}}}

\newcommand{\km}{\,\hbox{km}}

\newcommand{\mm}{\,\hbox{mm}}

\newcommand{\cm}{\,\hbox{cm}}

\newcommand{\lascia}[1]{}

\makeatletter
\def\hdashline#1(#2){\leavevmode\hbox to \z@{\baselineskip \z@%
\lineskip \z@%
\@dashdim=#2\unitlength%
\@dashcnt=\@dashdim \advance\@dashcnt 200
\@dashdim=#1\unitlength\divide\@dashcnt \@dashdim
\ifodd\@dashcnt\@dashdim=\z@%
\advance\@dashcnt \@ne \divide\@dashcnt \tw@ 
\else \divide\@dashdim \tw@ \divide\@dashcnt \tw@
\advance\@dashcnt \m@ne
\setbox\@dashbox=\hbox{\vrule \@height \@halfwidth \@depth \@halfwidth
\@width \@dashdim}\put(0,0){\copy\@dashbox}%
\put(#2,0){\hskip-\@dashdim\copy\@dashbox}%
\multiply\@dashdim 3 
\fi
\setbox\@dashbox=\hbox{\vrule \@height \@halfwidth \@depth \@halfwidth
\@width #1\unitlength\hskip #1\unitlength}\@tempcnta=0
\put(0,0){\hskip\@dashdim \@whilenum \@tempcnta <\@dashcnt
\do{\copy\@dashbox\advance\@tempcnta \@ne }}\@tempcnta=0
}\@makepicbox(#2,0)}

\def\vdashline#1(#2){\leavevmode\hbox to \z@{\baselineskip \z@%
\lineskip \z@%
\@dashdim=#2\unitlength%
\@dashcnt=\@dashdim \advance\@dashcnt 200
\@dashdim=#1\unitlength\divide\@dashcnt \@dashdim
\ifodd\@dashcnt \@dashdim=\z@%
\advance\@dashcnt \@ne \divide\@dashcnt \tw@
\else
\divide\@dashdim \tw@ \divide\@dashcnt \tw@
\advance\@dashcnt \m@ne
\setbox\@dashbox\hbox{\hskip -\@halfwidth
\vrule \@width \@wholewidth 
\@height \@dashdim}\put(0,0){\copy\@dashbox}%
\put(0,#2){\lower\@dashdim\copy\@dashbox}%
\multiply\@dashdim 3
\fi
\setbox\@dashbox\hbox{\vrule \@width \@wholewidth 
\@height #1\unitlength}\@tempcnta0
\put(0,0){\hskip -\@halfwidth \vbox{\@whilenum \@tempcnta < \@dashcnt
\do{\vskip #1\unitlength\copy\@dashbox\advance\@tempcnta \@ne }%
\vskip\@dashdim}}\@tempcnta0
}\@makepicbox(0,#2)}

\newdimen\sideftwd
\newbox\local@box\newbox\local@hbox

\def\lfigure#1{\if@twocolumn\local@ft{figure}{#1}\else\side@ft{figure}{#1}\fi}
\def\rfigure#1{\if@twocolumn\local@ft{figure}{#1}\else\side@ft{figure}{#1}\fi}
\def\ltable#1{\if@twocolumn\local@ft{table}{#1}\else\side@ft{table}{#1}\fi}
\def\rtable#1{\if@twocolumn\local@ft{table}{#1}\else\side@ft{table}{#1}\fi}
\def\endlfigure{\endlside@ft}
\def\endrfigure{\endrside@ft}
\def\endltable{\endlside@ft}
\def\endrtable{\endrside@ft}
\def\side@ft#1#2{\par
\sideftwd=#2
\def\@captype{#1}
\setbox\@tempboxa\vtop\bgroup\textwidth=\sideftwd
\columnwidth=\sideftwd \hsize\columnwidth
\@parboxrestore}
\def\endlside@ft{\egroup
\@tempdima=\ht\@tempboxa
\advance\@tempdima by \dp\@tempboxa
\@tempcnta=\@tempdima
\divide\@tempcnta by \baselineskip
\advance\@tempcnta by 2
\global\hangindent\sideftwd
\global\hangafter-\@tempcnta
\noindent \dp\@tempboxa=\z@ \ht\@tempboxa=\z@
\hbox to \z@{
\hbox to \z@{\hss\box\@tempboxa}\hss}%
\hskip\parindent
\global\@ignoretrue}
\def\endrside@ft{\egroup
\@tempdima=\ht\@tempboxa
\advance\@tempdima by \dp\@tempboxa
\@tempcnta=\@tempdima
\divide\@tempcnta by \baselineskip
\advance\@tempcnta by 2
\global\hangindent-\sideftwd
\global\hangafter-\@tempcnta
\noindent \dp\@tempboxa=\z@ \ht\@tempboxa=\z@
\hbox to \z@{\hskip\textwidth
\hbox to \z@{\hss\box\@tempboxa}\hss}%
\hskip\parindent
\global\@ignoretrue}
\def\local@ft#1{\def\@captype{#1}
\setbox\local@box\vbox\bgroup
\boxmaxdepth\z@\hsize0.9\columnwidth}
\def\endlocal@ft{\egroup
\[\hbox{\lower1ex\box\local@box}\]
\global\@ignoretrue}
\def\localfigure{\local@ft{figure}}
\def\localtable{\local@ft{table}}
\def\endlocalfigure{\endlocal@ft}
\def\endlocaltable{\endlocal@ft}
\makeatother
\newcount\Mac  \Mac=0
\def\putps(#1,#2)(#3,#4)#5#6{\ifnum\Mac=1 \put(#1,#2){\special{picture #5}}
\else  \put(#3,#4){\includegraphics{#6}} \fi}
\begin{document}
%\twocolumn[
\centerline{\bf 20/02/2000 \hfill    IFUP--TH/2000--00}
\centerline{\bf hep-ph/0002199 \hfill SNS--PH/00--04} \vspace{1cm}
\centerline{\LARGE\bf\Red Neutrino oscillations and large extra dimensions}\vspace{0.2cm}
\bigskip\bigskip\Black
\centerline{\large\bf Riccardo Barbieri$^{\rm a}$, Paolo Creminelli$^{\rm a}$ {\rm and} Alessandro Strumia$^{\rm b}$}
\bigskip
\centerline{(a) \em Scuola Normale Superiore, Piazza dei Cavalieri 7 and INFN, I-56126 Pisa, Italia}
\centerline{(b) \em Dipartimento di Fisica dell'Universit\`a di Pisa and INFN, I-56127 Pisa, Italia}

\bigskip\bigskip\Blue

\centerline{\large\bf Abstract}
\begin{quote}\large\indent
Assuming that right-handed neutrinos exist and propagate in some large extra dimensions, we attempt to give 
a comprehensive description of the phenomenology of neutrino oscillations. A few alternative explanations of
the atmospheric neutrino anomaly emerge, different from the standard $\nu_\mu \rightarrow \nu_\tau$ or $\nu_\mu 
\rightarrow \nu_{\rm sterile}$ interpretations. Constraints from nucleosynthesis and supernova 1987a
are discussed.
The constraints from SN1987a indicates a maximum radius of any extra dimension of about 1 $\AA$.
\end{quote}
\Black\vspace{1cm}

\section{Introduction}
The hypothesis that Standard Model (SM) singlets propagate in some extra dimensions
with relative large radii leads to striking consequences.
Most notable among them, if applied to the obvious candidate,
the graviton, is the possible disentanglement of the Planck scale from the scale where gravity
becomes strong~\cite{TeVgravity}.
It seems possible, in fact, that more surprises from the intense theoretical activity
on this and related subjects have yet to come.

In spite of this last remark, one can start asking if it will ever be possible to make any experimental observation 
related to these
phenomena. Always insisting on the graviton case, this has been and is being discussed both in the case of
particle collisions at high energy~\cite{ColliderGravitonico} and of tests 
of classical gravity~\cite{mmGravExp}.
With the relevant parameters at the boundary of the allowed region, such observations,
although not easy, may not be impossible.

After the graviton, the most natural candidate to propagate in some large extra dimension
is the right handed neutrino.
Interestingly enough, the smallness of the neutrino masses, of Dirac type, could in fact be a manifestation of this
very hypothesis~\cite{nuR5dA,DDG,nuR5d,ML,altro}.
The purpose of this paper is to make a first tentative exploration of the related experimental consequences.
A specific suggestion along these lines has already been made~\cite{DS}.

The plan of the paper is as follows. In section 2 we define the framework, 
including the flavour aspects. The possible connection with gravity is briefly 
outlined in section 3. In section 4 we diagonalize the neutrino mass
matrix for arbitrary values of the compactification radius in presence of
matter effects.
In section 5 we describe the oscillation amplitude of an interacting neutrino
with its 
Kaluza-Klein tower. In section 6 we show how the phenomenology of neutrino
oscillations gets modified. Possible signatures in on-going or future
experiments are summarized in section 7. In section 8 we show that big-bang 
nucleosynthesis is not manifestly inconsistent with the outlined phenomenology.
%%but could give,  after a careful and nontrivial study, the most significant constraints.
A strong constraint appears to come from the consideration of the neutrino luminosity
of the supernova 1987a, as discussed in section~9.
Such constraint indicates an upper limit on the largest radius of any extra dimension close to $1\AA$, which,
if taken at face value, would make impossible any observation of
oscillation phenomena discussed in this paper.
Our conclusions are summarized in section 10. 

We view this work as a contribution to the present discussion about possibilities of physics beyond the SM,
alternative to supersymmetric unification, which we still consider as the relatively more likely
description of nature at small distances.

\section{The framework defined}
Effectively, the action which defines our framework involves 5-dimensional massless
fermions $\Psi_i(x_\mu,y)$, the `right-handed neutrinos', one per generation, interacting on our brane
with the standard left-handed neutrinos $\nu_i$ in a way that conserves lepton number.
The relevant action is
\begin{equation}
\label{eq:5action}
S=
\int d^4x~dy~[\bar{\Psi}_i \Gamma_A i\partial^A \Psi_i] + 
\int d^4x~[\bar{\nu}_i i\ds \,\nu_i + \nu_i \lambda_{ij}\psi_j (x_\mu,0) h  +\hc]
\end{equation}
where $A=\{0,\ldots,4\}$, $\psi_j$ is one of the two independent Weyl spinors that compose
$\Psi_j=(\psi_j,\psi^c_j)$, $h$ is a normal Higgs doublet in four dimensions and $\lambda$ is a matrix
of Yukawa couplings with dimension of ${\rm (mass)}^{-1/2}$. As manifest from (\ref{eq:5action}), $\lambda$ can be made 
diagonal without loss of generality at the only price of introducing the usual unitary matrix $V$ which describes 
flavour changes in the charged current neutrino interactions~\cite{ML}. Redefining the $\nu_i$ in terms of the neutrino 
flavour eigenstates 
\begin{equation}
\label{eq:MNS}
\nu_\alpha^{(f)} = V_{\alpha i} \nu_i,\qquad \alpha= e, \mu, \tau, 
\end{equation}
after compactification of the 5-th dimension with suitable boundary conditions, the action (\ref{eq:5action}) becomes
$S = \int d^4x \sum_{i=1}^3 {\cal{L}}_i$, with   
\begin{equation}
\label{eq:4action}
{\cal{L}}_i = \bar{\nu_i} i \ds \nu_i + 
\sum_{n=-\infty}^{+\infty} \bigg[ \bar{\psi}_i^n i\ds \psi_i^n +
\bar{\psi}_i^{c\, n} i\ds \psi_i^{c\, n} + 
 m_i \,\nu_i\psi_i^n \bigg]+
\sum_{n=1}^{+\infty} \frac{n}{R} (\psi_i^n \psi_i^{c\, n} - \psi_i^{-n}  \psi_i^{c\, -n}) + \hc,
\end{equation}
where $R$ is the compactification radius
and $m_i$ are proportional to the eigenvalues of the Yukawa matrix $\lambda$.
Within this framework, $R$ is the only extra parameter that will
be introduced in the neutrino phenomenology other than the 3 masses $m_i$ and the mixing matrix $V$. 

As shown in appendix~\ref{AppA}, 
the Dirac nature of the mass matrix in (\ref{eq:4action}) is made explicit by an appropriate field redefinition, leading 
to the mass Lagrangian
$%  \begin{equation}\label{eq:Dirac}
  {\cal{L}}_{\rm mass} = \sum_{i=1}^3 N_{L i}^T M_i N_{Ri} + \hc
$ %  \end{equation}
with
  \begin{equation}\label{eq:matr}
  N_{Li} = \pmatrix{\nu \cr \nu_{nL} }_i,\qquad N_{Ri} =
  \pmatrix{\nu_R \cr \nu_{nR}}_i, \qquad n \ge 1
  \qquad\hbox{and}\qquad
  M_i = \pmatrix{m_i & \sqrt{2} m_i & \sqrt{2} m_i & \cdots \cr 0 & 1/R & 0
  & \cdots \cr 0 & 0 & 2/R & \cdots \cr
  \vdots & \vdots & \vdots & \ddots}.
  \end{equation}
To describe the evolution of a neutrino state of energy $E_\nu$, the relevant Hamiltonian is (one for any index $i$)
\begin{equation}
\label{eq:H}
{\cal{H}}^{(i)} = \frac{1}{2E_\nu} M_i M_i^T
\end{equation}
diagonalized by a matrix $U^{(i)}$ as
$U^{(i)} {\cal{H}}^{(i)} U^{(i)\,T} = {\cal{H}}_{\rm diag}^{(i)}.$
The elements $U_{0n}^{(i)}$ give the composition of the $\nu_i$ state in terms of the mass eigenstates corresponding
to the eigenvalues $\lambda_n^{(i)2}/2E_\nu R^2$ of ${\cal{H}}^{(i)}$. 

Postponing the explicit calculation of $\lambda_n^2$ and $U_{0n}$, the
amplitude 
$
A_{\alpha\beta}(t) = \langle \nu_\beta^{(f)} | \nu_\alpha^{(f)}(t) \rangle
$
for finding at any time $t$ a neutrino with flavour $\beta$, born at $t=0$ with flavour $\alpha$, is, using
(\ref{eq:MNS}),
\begin{eqnsystem}{sys:A}
\label{eq:Aab}
A_{\alpha\beta}(t) &=& \sum_{i=1,2,3} V_{\alpha i} V_{\beta i}^* A_i(t)\\
\riga{where}\\[-3mm]
\label{eq:Ai}
A_i(t) &=& A(\nu_i \rightarrow \nu_i,t) =\sum_{n=0}^{\infty} U_{0n}^{(i)2} \exp[i \lambda_n^{(i)2} t/2E_\nu R^2]. 
\end{eqnsystem}
Note the formal identity of (\ref{eq:Aab}) to the standard oscillation amplitude with the replacement 
$A_i(t) \rightarrow {\rm exp}(i m_i^2 t/2E_\nu)$.

Quite clearly, the neutrino phenomenology from~(\ref{sys:A}) can be significantly different from the standard one,
but only if $1/R$ is not too much bigger than all the $m_i$. Otherwise, for $1/R \gg m_i$, all of the Kaluza-Klein
(KK) states $\psi_i^n$, $\psi_i^{-n}$ with $n \ge 1$ in (\ref{eq:4action}) decouple and one remains with the standard 
Dirac masses $m_i \nu_i \nu_R^{i}$ only. In turn, if we want to influence the atmospheric or the solar 
neutrino anomalies, $1/R$ should not exceed $1\eV$ or so, {\em i.e.} $R \circa{>}  0.1\mu{\rm m}$.

\section{Connection with gravity}
Before going further, let us discuss the possible connection with 
gravity~\cite{DS}. Although not necessary, we make the simplifying
assumption that the graviton propagates in the same extra-dimensional space as the right handed neutrinos. Furthermore
the Planck mass $M_{\rm Pl}$ is related
to the fundamental scale of the theory $M_f$ and to the volume $V$ of the $\delta$ compactified extra dimensions
as $M_{\rm Pl} = M_f (M_f^\delta\:V)^{1/2}$.
To maintain a simple connection with gravity, we are in fact lead to consider several extra dimensions of different radii
$R_i$, with $R_{i > 1} \ll R_1$, so that (\ref{eq:4action}) is only an approximation where we neglect
the heavier KK excitations in the extra dimensions with small radii 
$R_{i>1}$~\cite{DS}.
%KK states  of higher mass
%\begin{equation}
%m_{n_1n_2 \cdots n_N} = \left(\sum_i \frac{n_i^2}{R_i}\right)^{1/2}
%\end{equation}
%with some $n_{i>1} \neq 0$.
All this is a necessity if $R_1\circa{>} 0.1 \mu{\rm m}$, as explained, and, for the heaviest
neutrino mass,
$m_{\rm max} \simeq {v}/(M_f^\delta V)^{1/2}$,
as it follows from a simple generalization of (\ref{eq:5action}) to $\delta$ 
extra dimensions~\cite{nuR5d}.
In this case
\begin{equation}
m_{\rm max} \simeq v \frac{M_f}{M_{\rm Pl}} \simeq 0.1 \eV \frac{M_f}{10^3 \TeV}
\qquad\hbox{and}\qquad
V \simeq \prod_i R_i \simeq \left(\frac{0.1 \eV}{m_{\rm max}}\right)^{\delta+2} 10^{26-15\delta} \frac{1}{\eV^\delta} 
\end{equation}
which shows, for any $\delta\ge 2$, the asymmetry of the different radii.

\section{Diagonalization of the neutrino Hamiltonian}\label{diagH}
Making reference for the details to appendix~\ref{AppA},
we describe in this section the diagonalization of the Hamiltonian\eq{H},
extended to include matter effects~\cite{MSW}, essential at least to discuss 
nucleosynthesis. It is
\begin{equation}\label{eq:Hmatter}
{\cal H}_{{\rm matter}}^{(i)}=\frac{1}{2E_\nu}[M_i M_i^T +\diag( {\mu_i\over R^2} ,0,0,\ldots)]
\end{equation}
where $\mu_i/R^2=2E^2_\nu(n_i^\pm-1)$ depends on the refraction indices $n_i^\pm$ for $\nu_i$ ($n_i^+$)
and $\bar{\nu}_i$ ($n_i^-$) in the relevant medium.
To be precise, the factorization of flavour as in eq.s~(\ref{sys:A}) is no longer exact
in presence of matter effects.
Since we will only study flavour mixing between $\mu$ and $\tau$ neutrinos
in media that do not distinguish $\mu$ from $\tau$ flavour, we can ignore this complication in the present work.

Still calling $\lambda_n^{(i)2}$ the eigenvalues of $2E_\nu R^2{\cal H}_{\rm matter}^{(i)}$, they satisfy the
eigenvalue equation (see appendix~A)
\begin{equation}\label{eq:lambda}
\lambda^2-\mu-\pi\lambda\xi^2\cot\pi\lambda=0
\end{equation}
where \Blue$\xi\equiv mR\Black$ and the index $i$ is left understood. Hereafter $R$ is set to unity unless explicitly reintroduced.
At the same time the matrix elements $U_{0n}$ are given by
\begin{equation}\label{eq:UU0n}
U_{0n}^2=\frac{2}{1+\pi^2\xi^2+\mu/\lambda_n^2+(\lambda_n-\mu/\lambda_n)^2/\xi^2}.
\end{equation}
Both\eq{lambda} and\eq{UU0n} extend the results of~\cite{DDG} for $\mu=0$. Note that,
for negative $\mu$, one $\lambda^2$ eigenvalue can be negative, so that the corresponding $\lambda$ is imaginary.
A qualitative description of $\lambda_n$ and $U_{0n}$ is as follows, depending on the sign of $\mu$.

\paragraph{$\mu>0$ case.}
For $\xi\to 0$ the eigenvalues $\lambda$ tend to the positive integers,
except for a special eigenvalue at $\lambda^2=\mu$, whose
eigenvector tends to coincide with the $\nu$-state up to mixings of order $\xi/n$.
When $\mu$ varies with the medium density or temperature and crosses a positive integer,
a resonant MSW conversion can take place.
For $\xi\gg 1$ the $\nu$-state is spread over a larger number of levels,
mostly those with $\lambda\approx \sqrt{\mu}\pm \pi\xi^2$,
each with a small mixing $U_{0n}^2\approx 1/\pi^2\xi^2$
and eigenvalues close to semi-integers.
Matter effects are not negligible if $\mu\circa{>}(\pi\xi^2)^2$
and do not suppress oscillations when $\mu$ is large.

\paragraph{$\mu<0$ case.}
The eigenvalues are as before, except for the special eigenvalue related to $\mu$.
For $\mu<-\xi^2$ the special eigenvalue $\lambda_0^2$ becomes negative and tends to $\mu$ as $\mu\ll-\xi^2$.
In this limit, the corresponding eigenstate is an almost pure $\nu$.
This means that, for large negative $\mu$, there is matter suppression.
By carefully expanding in $\xi^2/\sqrt{-\mu}$, for the component of the $\nu$-state on the special
eigenstate, one finds $U_{00}^2\simeq 1-\pi\xi^2/2\sqrt{-\mu}$.

\begin{figure}
\setlength{\unitlength}{1cm}
$$\begin{picture}(17,5)
\putps(0,0)(0,0){fPKK}{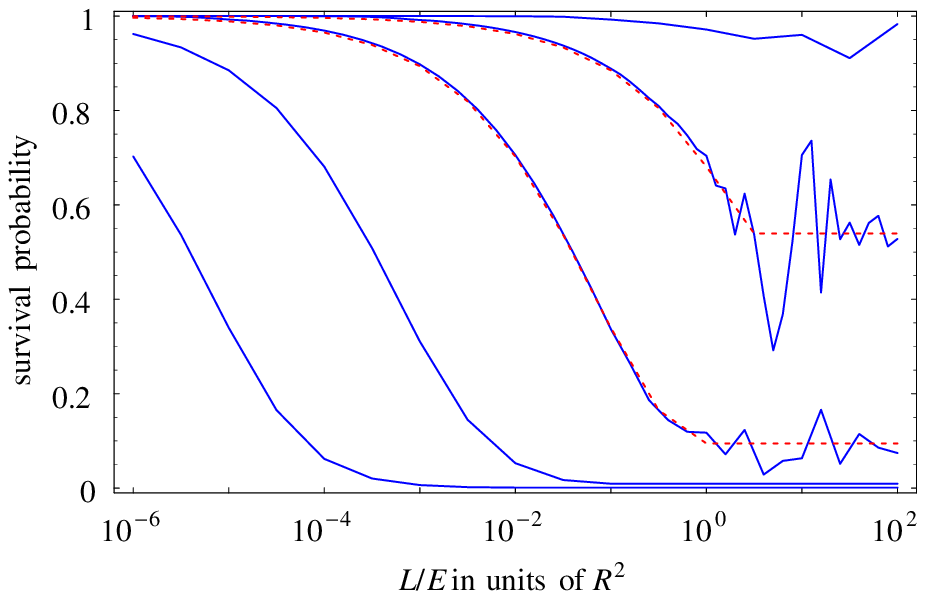}
\putps(10.5,0)(10.5,0){fPxi}{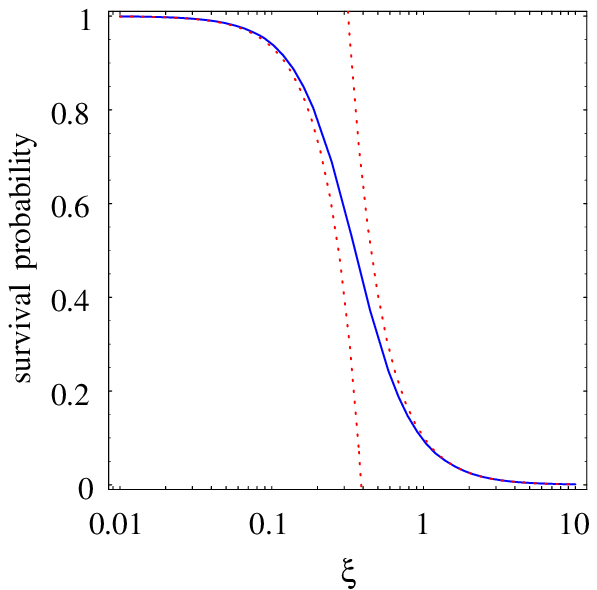}
\put(5,6.3){fig.\fig{P}a}
\put(14,6.3){fig.\fig{P}b}\Blue
\put(3.4,2){$\xi=10$}
\put(7.5,2){$\xi=1$}\Black
\end{picture}$$
\caption[1]{\em Oscillation probabilities.
Fig.\fig{P}a: $P_{\nu\nu}$ as function of $L/E_\nu R^2$ for $\xi =1/10,1/3,1,3,10$ (continuous
lines). For $\xi=1,1/3$ we also show the big $\xi$ approximation (dotted lines).
Fig.\fig{P}b: $P_{\nu\nu}$ as function of $\xi$ for $L\gg E_\nu R^2$ (continuous line)
and the small $\xi$ and big $\xi$ approximations (red dotted lines).\label{fig:P}}
\end{figure}

\section{Oscillation amplitudes for $\xi\circa{>}1$}
For small $\xi$, i.e. $1/R\gg m$, only few of the KK states mix with the standard neutrino and, furthermore,
with small mixing, of order $\xi$.
In this case it is not difficult to obtain the neutrino oscillation amplitudes.

Let us consider on the contrary the large $\xi$ limit, when the standard neutrino mixes with about $\pi\xi^2$ KK
states.
Matter effects are neglected for the time being.
For large $\xi$, $\lambda_n\approx n$ and we can safely approximate the sum in\eq{Ai} with an integral.
Using also, from\eq{UU0n}, $U_{0n}^2\approx 2/(\pi^2\xi^2+n^2/\xi^2)$ we have
\begin{equation}\label{eq:A(t)}
A \approx \int_0^\infty\!\! dn~ U_{0n}^2 e^{-n^2 (-iL/2E_\nu R^2)}=e^{z}(1-\erf\sqrt{z})\qquad
\hbox{where}\qquad
z=-i  \frac{L}{2E_\nu}\Big({\pi \xi^2\over R}\Big)^2.
\end{equation}
Note that the oscillation amplitude, obtained in the large $\xi$ limit,
depends only on the combination $\xi^2/R=m^2 R$ rather than on $m$ and $R$ independently.
For the probability $P_{\nu\nu}=|A|^2$ one has at large $L/E_\nu$
\begin{equation}\label{eq:Pinfty}
P_{\nu\nu}(L\to \infty) \approx \int_0^\infty\!\! dn~ U_{0n}^4 =  \frac{1}{\pi^2\xi^2}.
\end{equation}
A good analytic approximation for $P_{\nu\nu}$ at any $L$ is
$$P_{\nu\nu}\approx\max\bigg[\Big|1-\erf {\pi \xi^2\over R}\sqrt{-i  \frac{L}{2E_\nu}}\Big|^2,\frac{1}{\pi^2\xi^2}\bigg].$$
Fig.\fig{P} shows $P_{\nu\nu}$ as function of $L/E_\nu$ in units of $R^2$ for different values of $\xi$.
The comparison of the numerical result with the analytic approximation at $\xi=1,1/3$ shows the validity
of the analytic expression even at moderate values of $\xi$.
When $\xi$ gets sizable, the disappearance of the standard neutrino into the tower of KK states
becomes increasingly important. Matter effects on the oscillation amplitudes can be discussed along similar lines
(see appendix~A).

\begin{table}[t]
\begin{center}
\begin{tabular}{cccll} 
%\multicolumn{4}{c}{List of possibilities}\\ \hline \hline
case& $\xi_3$ & $1/R$ & solar & atmospheric  \\ \hline \hline \vspace{1mm}
A
& $\gg 1/3$
& \parbox{2cm}{$\ll3~10^{-3}\eV$\\  $\approx 3~10^{-3}\eV$\\  $\gg3~10^{-3}\eV$}
& \parbox{3cm}{only VO\\    SAM $\nu_e\to \nu_{\rm KK}$\\   standard}
& \parbox{4cm}{$\nu_\mu\to \nu_\tau,\nu_{\rm KK}$\\ with $\xi^2/R\approx 0.01\eV$.}\\ \hline 
B
& $\approx 1/3$
& $\approx 10^{-1}\eV$
& standard
& \parbox{4cm}{$\nu_\mu\to \nu_\tau,\nu_{\rm KK}^{\phantom{2}}$ or\\ $\nu_\mu\to \nu_{{\rm KK}_{\phantom{2}}}$}\\[1mm]   \hline
C
& $\ll1/3$
& $\circa{>} 10^{-1}\eV$
& standard
& standard $\nu_\mu\to \nu_{\tau}$\\[1mm] \hline\hline 
\end{tabular}
\caption{\em List of possibile oscillation patterns.
\label{tab:lista}}
\end{center}
\end{table}

\section{Unconventional fits of atmospheric and solar neutrino anomalies}
In this section we look for a comprehensive description of the neutrino anomalies,
both atmospheric and solar.
The new scale $1/R$ of KK excitations allows in many different ways to have
three different neutrino squared-mass splittings
that could, in principle, be associated with solar, atmospheric and LSND oscillations.
However, as in the standard phenomenology based on 3 neutrinos, we have not been able to also account for the LSND result.
%result without conflicting with SuperKamiokande,
%CHOOZ, Bugey, CDHS, CHORUS or NOMAD bounds,
%even allowing for a non standard solar fit.
A summary of the alternative possibilities that we have found is given in table\tab{lista}, as we now discuss.
They are characterized by the different ranges of $1/R$.
To keep things simple, we consider a hierarchical neutrino spectrum, with $m_3 > m_2 > m_1$.
Different spectra are only possible if all $\xi$ are small.

\subsection{Atmospheric neutrinos}
When $1/R$ decreases from about $1\eV$ the standard picture of neutrino oscillation is progressively perturbed.
In eq.s~(\ref{sys:A}) only $A_3$ is modified from the conventional form.
An increasing portion of the atmospheric neutrinos starts oscillating into their KK towers at the expenses
of the conventional $\nu_\mu\to \nu_\tau$ transition (case C in table\tab{lista}).
If $\xi_{1,2}\ll \xi_3 \ll 1$ approximate formulae for the transition probabilities in an intermediate $L/E_\nu$ range, {\em i.e.} 
$1/\Delta m^2_{ij}\gg L/E_\nu\gg R^2$, are
\begin{equation}
P_{\alpha\alpha} = 1 - \frac{2 \pi^2}{3} |V_{\alpha 3}|^2 \xi_3^2,\qquad
\hbox{and for $\alpha\neq\beta$}\qquad P_{\alpha\beta} = \frac{7\pi^4}{45}
|V_{\alpha 3} V_{\beta 3}|^2 \xi_3^4.
\end{equation}
Note that these formulae do not allow to interpret $1/R$ as the extra mass scale useful to account for the LSND
result. The situation in our case, with only one extra parameter $R$, is far more constrained than in the case 
where one adds one sterile neutrino with arbitrary mixing parameters.

\begin{figure}
\setlength{\unitlength}{1cm}
$$\begin{picture}(14,6)
\putps(7,0)(7,0){fCPfitMed}{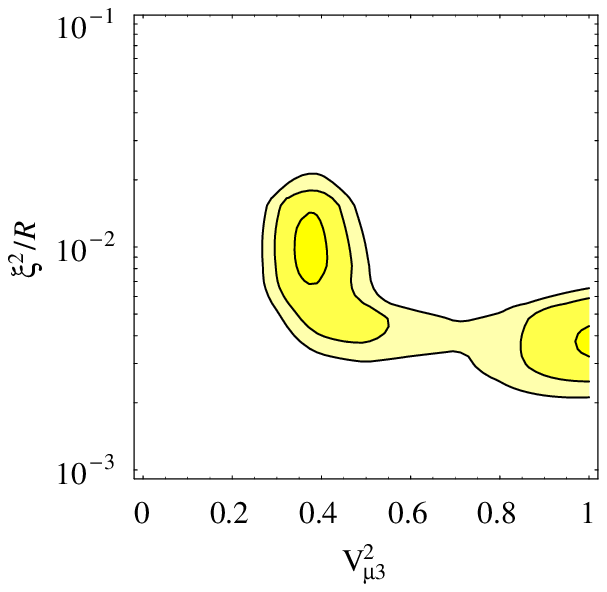}
\putps(0,0)(0,0){fCPfitBig}{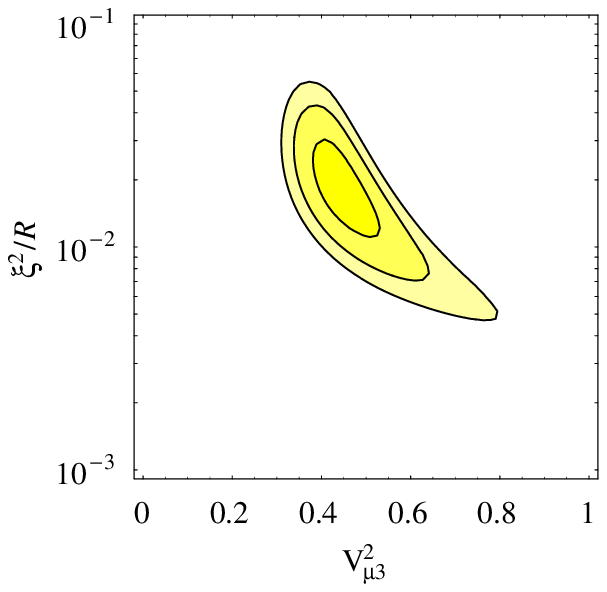}
\put(3,6.3){\fig{chiq}a: $\xi\gg1$}
\put(10,6.3){\fig{chiq}b: $\xi=1/2$}
\end{picture}$$
\caption[1]{\em fit of the SK atmospheric data in $(V_{\mu 3}^2,\xi^2/R)$.
Fig.\fig{chiq}a: $\xi\gg 1$ and any $R$.
Fig.\fig{chiq}b: $\xi=0.5$.
The contour lines correspond to $\chi^2=\{15,20,25\}$.
The best standard fit in terms of $\nu_\mu\to\nu_\tau$ gives $\min\chi^2\approx 14$.
\label{fig:chiq}}
\end{figure}

\begin{figure}[p]
\setlength{\unitlength}{1cm}
$$\begin{picture}(17,16)
%\putps(0,0)(0,0){fitBest}{fitBest.eps}
%\putps(0,6)(0,6){fitKK}{fitKK.eps}
%\putps(0,12)(0,12){fitBig}{fitBig.eps}
%\put(0,15){\fig{fit}a}
%\put(0,9 ){\fig{fit}b}
%\put(0,3 ){\fig{fit}c}
\putps(0,0)(0,0){ffits}{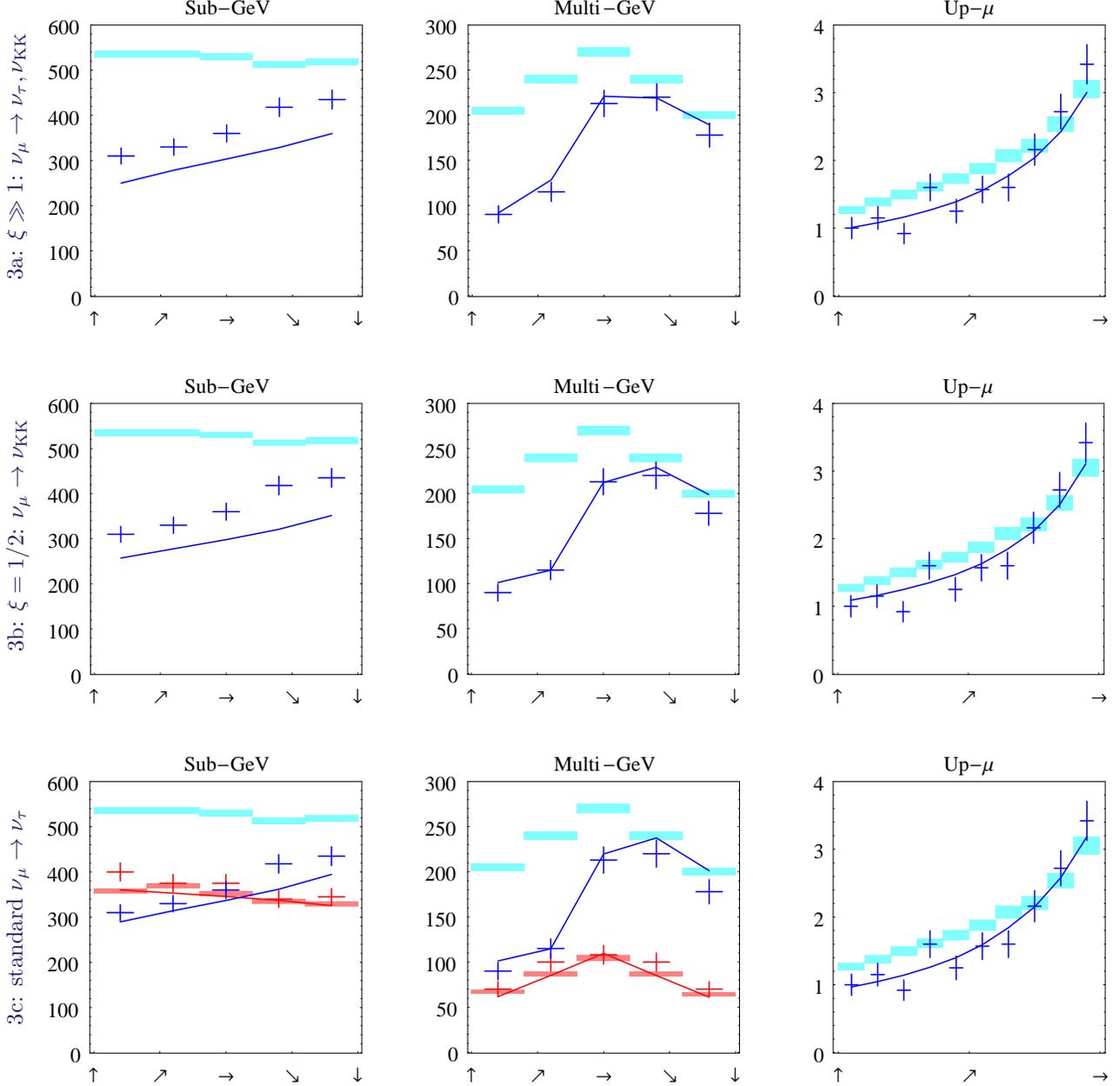}\Blue
\put(0,12.4 ){\begin{sideways}\fig{fit}a: $\xi\gg1$: $\nu_\mu\to\nu_\tau,\nu_{\rm KK}$\end{sideways}}
\put(0,6.7 ){\begin{sideways}\fig{fit}b: $\xi=1/2$: $\nu_\mu\to\nu_{\rm KK}$\end{sideways}}
\put(0,1){\begin{sideways}\fig{fit}c: standard $\nu_\mu\to\nu_\tau$\end{sideways}}\Black
\end{picture}$$
\caption[1]{\em Best fits of the SK zenith-angle distribution of
 sub-GeV (first column) and multi-GeV (second column)
$\nu_e,\nu_\mu$ events
and of upward-going muons (third column).
Each plot is drawn in the {\rm ($\cos\vartheta_{\rm zenith}$, number of events)} plane.
The arrows on the horizontal axes denote the direction of the scattered leptons.
Continuous lines denote fit predictions and
gray bars denote the no-oscillation predictions.
$\nu_\mu$ data are plotted in blue,
$\nu_e$ data are plotted in red and only in the last row,
since they are the same in all rows.
Crosses denote experimental data, including only statistical uncertainties.
The systematic uncertainty in the overall number of neutrino events in each sample of events
has been used to optimize the visual appearance of the standard fit in fig.\fig{fit}c:
as usual the most significant data are the shapes of the individual zenith-angle distributions.
Fig.\fig{fit}a: fit with $\xi\gg1$ ($\nu_\mu\to\nu_{\rm KK},\nu_\tau$: $\xi^2/R=0.015\eV$ and $V_{\mu3}^2=1/2$).
Fig.\fig{fit}b: fit with $\xi=1/2$ ($\nu_\mu\to\nu_{\rm KK}$: $\xi^2/R=0.004\eV$ and $V_{\mu3}^2=1$).
Fig.\fig{fit}c: standard fit ($\nu_\mu\to\nu_\tau$: $\Delta m^2_{23}=3~10^{-3}\eV^2, \sin^22\theta_{23}=1$).
\label{fig:fit}}
\end{figure}

To understand what happens when $\xi_3$ increases, let us consider the case when $\xi_3$ gets large.
As shown in fig.\fig{P}, for $\xi_3\circa{>}1$ and for sufficiently large $L/E_\nu$
the survival probability of $\nu_3$ drops below $10\%$.
In absence of flavour mixing, this would certainly disagree with the SuperKamiokande (SK) results~\cite{SKatm}.
In eq.s~(\ref{sys:A}) with $\xi_1,\xi_2\ll 1$ and neglecting matter effects
$A_1$ and $A_2$ are accurately approximated by the
standard phase $A_i=\exp(im_i^2 L/2E_\nu)$, which reduce to 1 at the $L/E_\nu$ relevant to the atmospheric neutrinos.
On the other hand, for $\xi_3>1$ and above the SK threshold for $\nu_\mu$ disappearance, we can take $A_3=0$.
From\eq{Aab} and above the SK threshold, using the unitarity of the $V$ matrix, we have therefore
\begin{equation}
P_{\alpha\alpha}\simeq (1-|V_{\alpha 3}|^2)^2,\qquad
\hbox{and for $\alpha\neq\beta$}\qquad
P_{\alpha\beta}\simeq |V_{\alpha3}V_{\beta3}|^2.
\end{equation}
Since the data suggest $P_{\mu\mu}\approx 0.5$ and $P_{ee}\approx 1$, a fit may be possible for
$V_{e3}\approx 0$ and $|V_{\mu 3}^2|\approx 0.4$.
These considerations are confirmed by an explicit fit of the data\footnote{The fit of SK atmospheric data is done as in~\cite{SKfit}.
In particular we have used:
(a) The prediction of atmospheric $\nu$ fluxes of~\cite{fluxes};
(b) The energy spectra of the parent atmospheric neutrinos,
corresponding to the various classes of events measured at SK
%(sub-GeV and multi-GeV, $e$-like, $\mu$PC and $\mu$FC),
as given by the Monte Carlo simulation of the SuperKamiokande detector,
available at the www address
$\hbox{\tt www.awa.tohoku.ac.jp/\~{}etoh/atmnu/e$\!\!\!${\_}nu/index.html}$.
(c) The latest SK data (848 days of exposure), as extracted from~\cite{data};
(d) The $\chi^2$ function defined in~\cite{chiq},
using 30 bins of experimental data and including all various systematic uncertainties.
%The best standard fit gives  $\chi^2\approx 14$.
`Official SK fits' employ unpublished data with finer energy and zenith-angle subdivisions of the neutrino-induced events.
If systematic errors are sufficiently low, these data could discriminate
the non-standard $L/E_\nu$ dependence of the oscillation probability from the standard one.}, shown in fig.\fig{fit}a.
As mentioned in the previous section, for large $\xi$, the oscillation amplitudes depend on the combination $\xi^2/R$,
which becomes the only relevant parameter other than $V_{\mu 3}$, since we set $V_{e3}=0$.
The contour plot of the $\chi^2$ in these parameters is shown in fig.\fig{chiq}a.
The value of $\xi^2/R$ around $0.02\eV$ is selected by the shape of the angular distribution of the muon neutrino
events. The fit does not fix the value of $1/R$, which can vary below about $10^{-2}\eV$ since $\xi>1$,
i.e.\ $m_3>1/R$.
This is case A in table\tab{lista}.

We can now ask what happens when $\xi_3$ is decreased below 1, in the intermediate
region between A and C.
Now $\xi_3$ and $1/R$, as well as the mixing angles, independently influence the fit.
In fig.\fig{fit}b we show the profile of the $\chi^2$ again in the plane $(\xi^2/R,|V_{\mu3}|^2)$
at fixed $\xi_3=1/2$.
%This value is picked up because at $\xi=1/2$ the individual oscillation probability $P_{\nu\nu}$ at large
%$L/E_\nu$ is close to $1/2$: about half of the interacting neutrinos oscillate into the KK states.
At this intermediate value of $\xi_3$  about half of the interacting neutrinos oscillate into the KK states.
As a consequence a fit of the atmospheric data is possible with $V_{\mu 3}\approx 1$, as illustrated in
fig.\fig{fit}b.
Unlike the case of very large $\xi$, or of the standard fit with $\xi=0$,
shown in fig.\fig{fit}c, no $\nu_\mu\to\nu_\tau$ oscillation is present in this case.

Back to fig.\fig{chiq}b, the two minima of the $\chi^2$ distribution are clear.
We have just discussed the case $V_{\mu3}\approx 1$. The other minimum originates from the one encountered
before at large $\xi$.
This should give an idea of what happens at various intermediate values of $\xi$. In this case (B in table\tab{lista})
$1/R$ ranges around $10^{-2}\eV$.

This completes our discussion of atmospheric neutrinos.
The fit at $\xi=1/2$ includes earth matter effects, whereas the one for large $\xi$ does not because in this case it
is difficult to compute neutrino propagation across few layers with different density.
%From a preliminary investigation,
We expect that the main features of solution A will remain unchanged.

\subsection{Solar neutrinos}
Given these alternative descriptions of the atmospheric neutrino anomaly, we should ask now how the
solar neutrino deficit can be accounted for.
So far $m_2$ and $m_1$ have not been fixed.
We only required that $\xi_2$ and $\xi_1$ are both small.  This being the case, it is in fact simple to see
that a standard description of solar neutrinos is possible in all cases
with negligible interference of the KK towers.
For small values of $1/R$ there is however the possibility of a transition $\nu_e\to\nu_{\rm KK}$
using the MSW effect, which is compatible with the solar data~\cite{DS}.
It requires $1/R\approx 3~10^{-3}\eV$ and a mixing with the KK states
determined by $\xi_2\approx 0.01$, or $m_2\approx 10^{-(4\div 5)}\eV$,
so that a fit of SK atmospheric data requires $\xi_3\sim 2$.
When the parameter $\mu$ of section~\ref{diagH}
is specified for the electron neutrino and with the solar density profile, the resonant MSW conversion
mentioned there ($\mu$ positive, small $\xi$) takes place and suppresses the different components
of the solar $\nu_e$ spectrum as possibly observed by the various solar neutrino experiments.

\begin{figure}[t]
\setlength{\unitlength}{1cm}
$$\begin{picture}(12,6)
\putps(0,-0.5)(0,-0.5){fSKoscs}{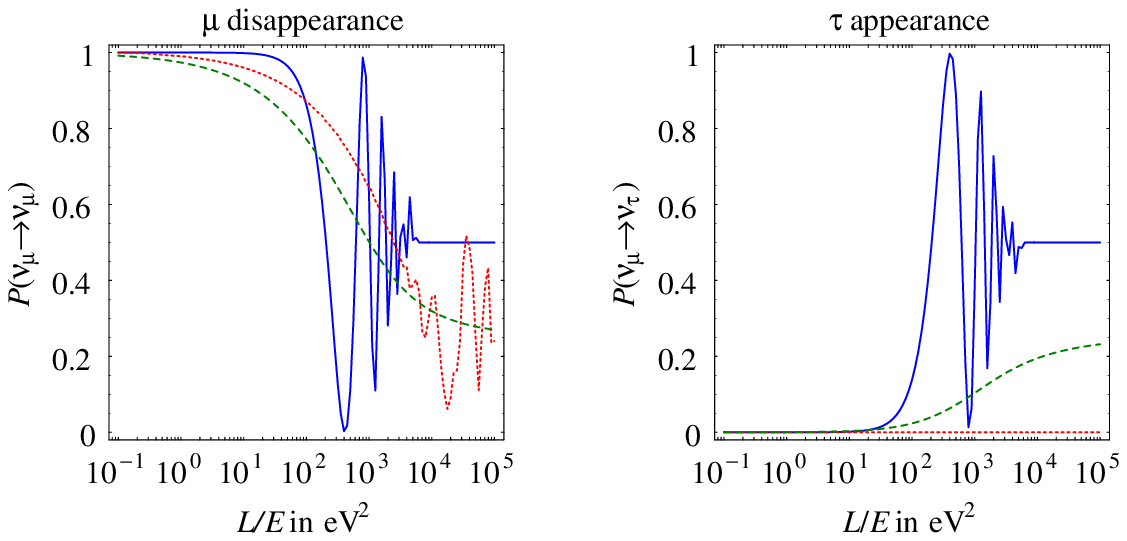}
%\Green\put(8,1){$P_{\mu\tau} \propto L$}
%\Blue\put(8.5,2){$P_{\mu\tau} \propto L^2$}\Black
\end{picture}$$
\caption[SKP]{\em The $P_{\mu\mu}$ (fig.\fig{SKP}a) and $P_{\mu\tau}$ (fig.\fig{SKP}b) that give the best SK fits.
Continuous blu line: standard $\nu_\mu\to \nu_\tau$ fit.
Dotted red line: $\nu_\mu\to \nu_{\rm KK}$ fit with intermediate $\xi=1/2$.
Dashed green line: $\nu_\mu\to \nu_\tau,\nu_{\rm KK}$ fit with large $\xi$.
\label{fig:SKP}}
\end{figure}

\section{Special features of the proposed solutions}
Some alternative descriptions of the atmospheric neutrinos appear possible. The crucial point, however, 
would be to indicate precise signatures of such solutions visible in appropriate
neutrino experiments.
%In the following we only give some preliminary considerations on this issue.
To this purpose fig.s\fig{SKP} are of interest.
We give there, versus $L/E_\nu$, the probabilities $P_{\mu\mu}$ and $P_{\mu\tau}$ that correspond to the fits of the SK results
shown in fig.\fig{fit}.
A few features of these plots might be relevant for an experimental discrimination of the
various possibilities.

\begin{enumerate}
\item The absence of a first clear dip in the $L/E_\nu$-shape of $P_{\mu\mu}$ is a characteristic
of the KK fits that we have discussed at intermediate and big $\xi$, at clear variance with the shape
of $P_{\mu\mu}$ in the standard $\nu_\mu\to\nu_\tau$ interpretation of the data.

\item
The non-standard transition from unoscillated to oscillated atmospheric neutrinos
requires a $L/E_\nu$-range longer than the standard one and even the
one that would be produced by neutrino decay~\cite{NuDecay}.
Therefore, unlike what happens in the standard case,
a good fit of atmospheric data significantly constrains the outcome of $\nu_\mu$ disappearance experiments.
For example the on-going K2K experiment~\cite{K2K} should observe only $65\%\div 85\%$ of the events with respect to the
no-oscillation case, while the larger range $30\%\div90\%$ is allowed  by a standard fit of SK data.
Hence K2K could put the constraint $R<0.01\,{\rm mm}$ on the scenario under study.

\item 
A related characteristic feature of the same $\nu_\mu$ survival probability is a precocious disappearance
of $\nu_\mu$ (and appearance of $\nu_\tau$ if $V_{\tau 3}\neq 0$) at relatively low $L/E_\nu$.
%%By comparing fig.s\fig{fit}a,b with fig.\fig{fit}c we see how the downward-going $\nu_\mu$ events at SK are affected.
The effect is not big since otherwise it would have made difficult the same SK fit or it could have
been in conflict with the null result of CHORUS and NOMAD~\cite{CHORUSeNOMAD}. At small $L/E_\nu$, an analytic approximation for
$P_{\alpha\beta}$ is
\begin{equation}
\label{largexi}
P_{\alpha\beta} \simeq 2\pi |V_{\alpha 3} V_{\beta 3}|^2  \Big(\frac{\xi_3^2}{R}\Big)^2 \frac{L}{E_\nu}.
\end{equation}
Note the linear dependence on $L$, rather than the quadratic one characteristic of the standard oscillation formula.
The effects related to (\ref{largexi}) are not without interest already for the downward going $\nu_\mu$ in
SK, in view of the parameters shown in fig.\fig{chiq}.

\item 
A partial suppression of the $\nu_\tau$-events might occur in an appearance experiment trying to measure
$P_{\mu\tau}$ at {\em large} $L/E_\nu$ relative to the expectation in the case of
the standard $\nu_\mu\to\nu_\tau$ interpretation of the data.

\item  Earth matter effects suppress the oscillations of a long-baseline beam of $\nu_\mu$ with large
energy $E_\nu\circa{>}10\GeV$, although with a slower $E_\nu$ dependence than in the usual case.
On the contrary earth matter effects enhance the oscillations of a $\bar{\nu}_\mu$ beam.

\end{enumerate}
All these points deserve further quantitative investigation.

\section{Constraints from nucleosynthesis}
So far we have compared the phenomenology of neutrinos from extra dimensions with oscillation experiments.
%Constraints on this phenomenology may however arise from other sources, like lepton universality~\cite{lu}
%stellar cooling, the neutrino observation from the supernova 1987A or cosmology~\cite{DS}.
%We have looked in all these pieces of physics and we have not found any obvious major constraint that would
%exclude the parameters discussed in the previous sections in a clear way.
%A special problem is represented by standard big-bang nucleosynthesis, as we now discuss.
In this section we study the constraint arising from standard big-bang nucleosynthesis.
As well known, the danger for nucleosynthesis is that too many KK neutrino modes are produced before the time of nucleosynthesis.
%altering in this way the equilibrium rate of the crucial nucleosynthesis reactions at temperatures at about $1\MeV$.
Two different production mechanisms have to be studied: (i) by incoherent scattering; (ii) by coherent oscillations.

In short, the incoherent production is not dangerous provided a temperature $T^*$ is assumed,
at which the abundance of relic KK neutrinos is negligibly small.
$T^*$ is in the GeV range for $\delta=2$ and can easily be much higher as $\delta$ increases.
This agrees qualitatively with several statements in the literature~\cite{nuR5d,DS}.

The case with coherent oscillations is more delicate since the evolution of an infinite number of different
neutrino mass eigenstates is not easy to follow, especially in the case of a sizable
$\xi$ parameter.
%We remind from section~\ref{diagH} that a sizable $\xi$ does not lead to a large mixing of any KK state
%with the interacting neutrino, but rather to a large number of states all with comparable small mixing.

\begin{floatingfigure}[r]{8cm}
\includegraphics{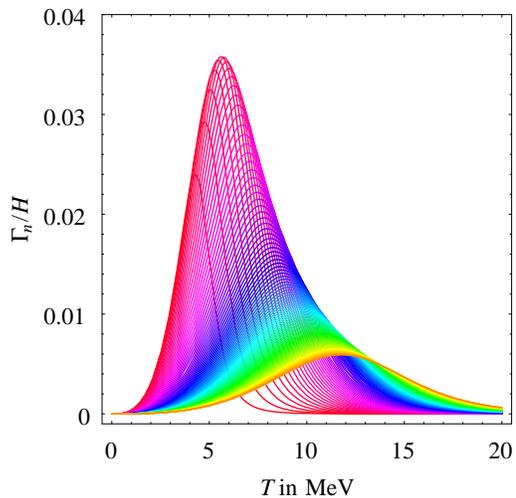}
\caption{\em Values of $\Gamma_n/H$ for $\xi=2$ and $1/R=2.5~10^{-3}\eV$
for $n=1,\ldots 100$.}\label{fig:modi}\vspace{3mm}
\end{floatingfigure}
A simple way to get an idea of what happens is the following. Since the oscillation frequencies are large relative
to the collision or the expansion rates at any temperature $T$ close to nucleosynthesis,
one can estimate if any one of the mass eigenstates may reach equilibrium by comparing to the expansion rate $H(T)$
the effective interaction rate of the $n$-th state
$\Gamma_n(T)\approx U_{0n}^2 \Gamma_\nu(T)$.
Here $\Gamma_\nu(T)$ is the typical interaction rate of a standard neutrino with the medium and
$U_{0n}^2$ is given in\eq{UU0n} in terms of the matter potential $\mu$.
If one neglects any matter asymmetry in the primordial medium, either original or generated by the neutrino evolution
equations themselves~\cite{mu}
\begin{equation}
\mu = -2 c (E_\nu R G_{\rm F})^2 T^4/ \alpha_{\rm em}
\end{equation}
where $\alpha_{\rm em}$ and $G_{\rm F}$ are the fine structure and Fermi constants, $E_\nu$ is the neutrino energy, to be suitably
averaged, and $c$ is a numerical coefficient close to $0.2$ for $\mu $ and $\tau$ neutrinos.
The $\mu$ parameter has a negative fixed sign
both for neutrinos and antineutrinos. As discussed in section~\ref{diagH}, a negative $\mu$ implies a MSW suppression
of $\Gamma_n(T)$. Fig.~\ref{fig:modi} shows the plot of $\Gamma_n(T)/H(T)$ versus
$T$ for $\xi = 3$ for a large number of mass eigenstates different from the special one, which coincides with
the interacting neutrino at large temperature.
The rise at small $T$ is a manifestation of the $T^3$-behavior of $\Gamma_\nu/H$, whereas
the fall  at large $T$ is a consequence of the MSW suppression of the effective mixing factor $U_{0n}^2$. None of the
states is close to equilibrium at any temperature. The opposite would have been a clear problem for
nucleosynthesis since the typical temperatures are close to the MeV.
It is harder to say if the situation described 
in fig.~\ref{fig:modi} is compatible with standard nucleosynthesis. Even though the energy density stored in any
state is small because none gets even close to equilibrium, their total energy density might be significant.

To answer this question requires solving the coupled system of infinite equations
for the neutrino density matrix $\rho$.
Here we only check that standard big-bang nucleosynthesis is not grossly inconsistent with the picture described above.
In the $T$-dependent ${\cal H}_{\rm matter}(T)$ eigenstate basis,
neutrino propagation averages to zero the off-diagonal elements of $\rho$
much faster than the typical interaction or expansion rates at $T\sim \MeV$.
Setting to zero the off-diagonal elements $\rho_{nm}$, $n\neq
m$, and assuming that neutrinos are in kinetic equilibrium,
the evolution equations for the neutrino densities $N_n\equiv \rho_{nn}$
of all individual neutrino
mass eigenstates, normalized to the equilibrium density, are
\begin{equation}
\frac{d N_n}{d\ln T} = -\frac{\Gamma_\nu(T)}{H(T)}U_{0n}^2(T)(1-N_n)+\cdots.
\end{equation}
We have explicitly included annihilation processes
and indicated with
$\cdots$ additional terms due elastic scatterings.
These terms have comparable rates and
redistribute the total neutrino number $N$ in its various components,
but do not directly affect the total neutrino number $N\equiv \sum_n N_n$.
Summing all the equations we find the evolution equation
for the total neutrino density
\begin{equation}\label{eq:dN}
\frac{d N}{d\ln T} = -\frac{\Gamma_\nu(T)}{H(T)}(1-N_\nu),
\end{equation}
where $N_\nu\equiv \sum_n U_{0n}^2 N_n$ is the number density of the interacting neutrino in the flavour basis.
%%\begin{eqnsystem}{sys:sumn}
%%\sum_n \Gamma_n &=& \Gamma_\nu\sum_n U_{0n}^2 = \Gamma_\nu\\
%%\riga{and}\\[-2mm]
%%\sum_n \Gamma_n N_n &=& \Gamma_\nu\sum_n U_{0n}^2N_n = \Gamma_\nu N_\nu
%%\end{eqnsystem}
If neutrinos start with a normalcy distribution at some temperature $T$,
oscillations rapidly convert it into $N_n(T)=U_{0n}^2(T)$
so that $N_\nu=\sum_n U_{0n}^4  = P_{\nu\nu}$.
Using this value of $N_\nu(T)$ and the approximation
$P_{\nu\nu}\approx U_{00}^4\approx 1-\pi\xi^2/\sqrt{-\mu}$ 
valid at large $\mu$, or large $T$, we estimate the right-hand side of\eq{dN} as
\begin{equation}\label{eq:dNappr}
\frac{d N}{d\ln T} \approx  - \frac{\Gamma_\nu(T)}{H(T)}\frac{\pi\xi^2}{\sqrt{-\mu(T)}}.
\end{equation}
As for the individual rates $\Gamma_n/H$, the right-hand side of\eq{dNappr} is small at $T\circa{<}1\MeV$
and progressively increases.
However, unlike $\Gamma_n/H$,
it does not get to a maximum but rather
flattens out to a horizontal asymptote at a level ${\cal N}(\xi^2/R)/(10^{-2}\eV)$.
In the present case where $\nu$ is a combination of $\nu_\mu$ and $\nu_\tau$,
the numerical coefficient ${\cal N}$ is of order unity and has a large uncertainty.
Therefore we cannot conclude that values of
$\xi^2/R\approx 10^{-2}\eV$, relevant to the atmospheric neutrino anomaly,
give neutrino oscillations
incompatible with standard big-bang nucleosynthesis.
A detailed and non trivial study could give significant constraints.
An unambiguous problem, if present, could not be avoided as usual by invoking a large lepton asymmetry which, in the case under study,
can only suppress $\nu$ {\em or} $\bar{\nu}$ oscillations, but not both.
%In our view, however, even if a conflict with standard big-bang nucleosynthesis were present,
%it would anyhow be worth pursuing a direct comparison with neutrino oscillation data.

\section{Constraints from supernova 1987a}
In this section we show that the SN1987a observation is likely to put a very severe constraint on $R$,
not compatible, if taken at face value, with the phenomenology of neutrino oscillations as discussed
in the previous sections.
This is because a large number of KK neutrinos would carry away from the supernova (SN) too much energy in invisible channels,
thus shortening the observed neutrino burst in an unacceptable way.
More precisely, we demand that the energy loss rate in invisible channels, $W_{\rm inv}$, is less
than about $10^{19}\,{\rm erg}/{\rm gr}\cdot\sec$ for typical average conditions of a SN core with a
density $\rho\approx 3\cdot 10^{14}{\rm gr}/\cm^3$ and a temperature $T\approx 30\MeV$~\cite{SN1987}.
As in the case of nucleosynthesis, one has to consider the production of KK neutrinos
both by incoherent scatterings and by coherent oscillations.

The rate of incoherent production of a single KK state is estimated as
$\Gamma_{\rm inc}\approx (m/E)^2\Gamma_\nu$,
where $m$ is the largest of the masses $m_i$ introduced in section~2,
$E\approx100\MeV$ is a typical neutrino energy and $\Gamma_\nu\approx G_{\rm F}^2n_N E^2$
is the collision rate of a standard neutrino in the supernova core, in terms of the nucleon
density $n_N$.
Accounting for the number of KK states, approximately $RE$, the corresponding energy loss rate can be estimated as
$W_{\rm inv}\approx (m/E)^2 ER \cdot W_\nu$, where $W_\nu\approx 10^{27}{\rm erg}/{\rm gr}\cdot\sec$ is the energy loss rate
produced by $\Gamma_\nu$ if the standard neutrinos were not trapped.
Requiring $W_{\rm inv}/W_\nu \circa{<}10^{-8}$ gives
$
(10^{-3}\eV\cdot R)({m^2}/{10^{-3}\eV^2})\circa{<}1
$
which is not incompatible with the values of $m$ and $R$ considered in the previous sections.
Note, however, that had we considered $\delta>1$ extra dimensions, all with comparable radii,
the increase in the number of KK states of energy less than $E$, about $(RE)^\delta$, 
would have lead to a stronger constraint on $R$
\begin{equation}\label{eq:R-coherent}
R<0.2\mm \cdot 10^{-11(1-1/\delta)} \left(\frac{m^2}{10^{-3}\eV^2}\right)^{-1/\delta}.
\end{equation}
Let us now consider the effect of coherent oscillations of the active neutrino state
associated with the mass $m$ into its KK tower.
Depending on the sign of $\mu$ in the supernova core, MSW resonant oscillations will take place
either in the neutrino or in the antineutrino channel. Since the neutrino effective potential $V$ in the core
is of order $10\eV$, the KK states interested in the resonant transitions have a mass $m_n\approx n/R\approx (VE)^{1/2}\approx
10^{4\div 5}\eV$, with a number of states interested in the resonant transition of order $n_{\rm res}\approx R \cdot 10^{4\div 5}\eV$.
The effect of these oscillations can be reliably estimated for small $\xi=mR$.
This is so because in such a case, due to the small vacuum mixing angle  between the $n$-th KK state and the standard
neutrino, $\theta_n\approx \xi/n$, the width of the $n$-th resonance is smaller than the separation between
two contiguous resonances.
As a consequence, the survival probability for a standard neutrino (or antineutrino) produced in the core can be computed
as the product of the survival probabilities in each resonance,
$P_{\nu\nu}\approx \prod_n P_n$~\cite{DS}.
Furthermore, for each $P_n$ one can take an approximate formula describing the transition
with initial and final densities respectively much larger and much smaller than the resonance density,
$P_n\approx e^{-\pi \gamma_n/2}$, where
$$\gamma_n \approx \left.\frac{4\xi^2}{R^2E}\frac{V}{dV/dr}\right|_{\rm res}$$
is the adiabaticity parameter at the $n$-th resonance crossing.
Approximating $V/(dV/dr)$ with the radius of the core, $r_{\rm core}\approx 10\km$, $\gamma_n$ is of order
$m^2/10^{-3}\eV^2$ and is approximately $n$-independent.
Hence the individual $P_n$ are expected to deviate sensibly from unity, which is a fortiori the case for
the overall probability $P_{\nu\nu}$ due to the large number of  resonances crossed.
This is untenable no matter what the fate will be of the KK states produced in the oscillations,
either escaping or trapped in the core, since in both cases the luminosity in standard neutrinos will be
drastically reduced.
We do not even see how such a conclusion could be avoided by having $1/R$ approaching $m$, or $\xi\circa{>}1$,
even though a precise calculation in this case is not easy to make due to the overlap of the different resonances.

We therefore believe that this sets an upper limit on $R$ close to $1/10^{4\div 5}\eV$.
Above this value no resonance is crossed anymore and the mixing angle of any KK state with the standard neutrinos becomes
negligibly small. This bound, together with\eq{R-coherent}, would seem to indicate that independently from the number of dimensions,
the largest of the radii cannot be bigger than about $1\AA$.

\section{Conclusions}
In the discussion about possible patterns of physics
beyond the SM, the
existence is being considered of some 
large extra dimensions where SM singlets could propagate. The possibility
of lowering the scale where gravity becomes 
strong is a striking consequence of this hypothesis. Hence, the hierarchy
problem is generally quoted as its main
motivation. On the other hand, unlike the case, e.g., of supersymmetric
unification, we know of no observation, direct
or indirect, that supports this picture. The problem of finding possible
signals of large extra dimensions is,
therefore, particularly acute.

In this paper we have shown that neutrinos offer a concrete possibility of
testing the existence of large extra dimensions,
alternative to those considered in high energy collisions or in gravity
measurements. This is so if there is a large   
extra-dimension that extends above $0.1 \mu {\rm m}$, where right-handed
neutrinos propagate. We think that neutrino oscillation
experiments can fully explore this range of distances.
Constraints arise in principle on this phenomenology both from standard big-bang nucleosynthesis
and from SN1987a.
As we have discussed, particularly constraining appears to be the consideration of the neutrino
luminosity from SN1987a.
If taken at face value, a maximum radius of any extra dimension is implied by these considerations
close to $1\AA$, making impossible any observation in neutrino oscillation phenomena.
This bound is likely to be valid, even though we cannot perform a clear calculation in all the
relevant range of the parameters.
However ad hoc ways to evade it without affecting the signals discussed in sections 6, 7
may exist, like suitable decays of the heavier KK neutrinos in the $100\,{\rm KeV}$ 	range, which
undergo the resonant transitions in the supernova core. In our view, it is anyhow worth pursuing a direct comparison
with neutrino oscillation data.

% We are not aware of any independent constraint capable of
% excluding in a clear way the range of parameters that we have considered.
% This includes also the consideration of standard 
% big-bang nucleosynthesis, to the best of our knowledge.
% 
% The simplest possibility is that the graviton propagates in the same
% extra-dimensional space as right-handed neutrinos.
% In this case, it is possible to keep one extra-dimension with a large
% enough radius, $R_1$, if the radii of the other 
% extra-dimensions are much smaller. Independently from neutrino physics,
% the asymmetry of the radii is a necessity also to 
% maintain $R_1$ in the range of sensitivity of the planned sub-millimeter
% gravitational measurements. In such a case, the
% theory with several extra dimensions can be tested both by classical
% gravity measurements and by neutrino oscillations, while
% avoiding astrophysical or other laboratory bounds. 

\paragraph{Acknowledgments}
Work supported in part by the E.C. under TMR contract No. ERBFMRX--CT96--0090.
We thank A.D.\ Dolgov and R. Rattazzi for useful discussions.

%\small
\begin{appendix}
\section{Explicit diagonalization of the Hamiltonian}\label{AppA}
From\eq{4action} we arrive to the Dirac mass matrix\eq{matr} with the
following definitions:
\begin{equation}
\nu_R^n = \frac{1}{\sqrt{2}} (\psi_i^n + \psi_i^{-n}) \qquad  \nu_L^n =
\frac{1}{\sqrt{2}} (\psi_i^{c\:n} - \psi_i^{c\:-n})
\quad {\rm for} \quad n\ge 1 \qquad{\rm and}\qquad \nu_R = \psi_i^0.
\end{equation}
In this way $\psi_i^{c\:0}$ and the linear combinations orthogonal to
$\nu_R^n$ and $\nu_L^n$ decouple and can be forgotten.
The resulting effective Hamiltonian\eq{Hmatter} for the first $k_{\rm max}$ KK states is
${\cal H}_{\rm eff} = X/2E_\nu R^2$ where
the dimensionless matrix $X$ has an `almost diagonal' form:
keeping only the first $k_{\rm max}$ KK excitations
its only non-vanishing off-diagonal elements are $X_{k0}=X_{0k}=k\sqrt{2}\xi$ ($k\ge 1$).
Its diagonal elements are $X_{kk}=k^2$ for $k\ge 1$ and $X_{00}=(2k_{\rm max}+1)\xi^2 + \mu$, where
the $\mu$ term parameterizes matter effects.
Calling $\lambda^2$ the eigenvalues of $X$,
the eigenvalue equation can be written in the closed form
$$
\det[X - \lambda^2\One]=
\bigg[(X_{00}-\lambda^2)-\sum_{k=1}^{k_{\rm max}} \frac{X_{k0}X_{0k}}{X_{kk}-\lambda^2}\bigg]
\prod_{k=1}^{k_{\rm max}}(X_{kk}-\lambda^2)
=
\bigg[\xi^2+\mu-\lambda^2-2\xi^2\sum_{k=1}^{k_{\rm max}}\frac{\lambda^2}{k^2-\lambda^2}\bigg]
\prod_{k=1}^{k_{\rm max}}(k^2-\lambda^2)
%\label{eq:secular}
$$
In the limit $k_{\rm max}\to\infty$ we obtain the eigenvalue equation\eq{lambda}.
%%%$\lambda^2-\mu-\pi\lambda\xi^2\cot\pi\lambda =0$.
As already remarked, when $\mu < 0$ the squared mass matrix may have
negative eigenvalues: for $\mu < \xi^2$, in fact,\eq{lambda}
admits one $\lambda^2 <0$ with $\lambda$ purely imaginary.

For any eigenvectors $e_i(\lambda)$ the secular equation gives a relation
between the various elements and the first one:
$e_k/e_0=-k\sqrt{2}\xi/(k^2-\lambda^2)$. Using the normalization condition
$\sum_i e_i e_i =1$ we arrive at
\begin{equation}
\frac{1}{U_{0n}^2} = \frac{1}{e_0^2(\lambda_n)}=1+2\xi^2
\sum_{k=1}^\infty\frac{k^2}{(k^2-\lambda^2_n)^2}=
\frac{1}{2}\bigg[2+\pi^2\xi^2(1+\cot^2\pi\lambda_n)-\frac{\pi^2\xi^2}{\lambda_n}\cot\pi\lambda_n\bigg],
\end{equation}
which, using\eq{lambda}, gives\eq{UU0n}.
It is not difficult to obtain simple expressions that
extend the large-$\xi$ oscillation amplitudes\eq{A(t)} and\eq{Pinfty} to include matter effects.

If $\mu<0$, one has to separately include the contribution of the special eigenstate with $\lambda^2<0$.
When $\mu\ll -(\pi\xi^2)^2$ this state has a small KK component and
$P_{\nu\nu}(L\to\infty)\approx U_{00}^4\approx 1-\pi\xi^2/\sqrt{-\mu}$ so that matter effects suppress oscillations.
On the contrary when $\mu>0$ eq.\eq{Pinfty} remains valid.
Moreover when the matter potential is very large,
$\rho \equiv \mu/(\pi\xi^2)^2\gg 1$,
the effective $L/E_\nu$ above which neutrino oscillations are significant
is {\em reduced\/} by a factor $\rho^{1/2}$ with respect to the $\mu=0$ value given by eq.\eq{A(t)}.

\end{appendix}

\footnotesize
\begin{multicols}{2}

\end{multicols}

\end{document}